\documentclass[doublecol,a4paper,showpacs]{epl2}
\usepackage{graphicx}
\usepackage{color}
\usepackage{amsmath}
\usepackage{amssymb}
\usepackage{enumerate}
\usepackage[colorlinks=true, pdfstartview=FitV, linkcolor=blue, citecolor=red, urlcolor=black, breaklinks=true]{hyperref}
\newcommand{\be}{\begin{equation}}
\newcommand{\ee}{\end{equation}}
\newcommand{\ben}{\begin{eqnarray}}
\newcommand{\een}{\end{eqnarray}}
\newcommand{\bes}{\begin{subequations}}
\newcommand{\ees}{\end{subequations}}

\newcommand{\vphi}{\varphi}
\begin{document}
\title{Quasi-Compact Q-balls}

\author{D. Bazeia\inst{1} \and M.A. Marques\inst{1} \and R. Menezes\inst{2,1}}
\shortauthor{D. Bazeia \etal}

\institute{                    
  \inst{1} Departamento de F\'\i sica, Universidade Federal da Para\'\i ba, 58051-970 Jo\~ao Pessoa, PB, Brazil\\
  \inst{2} Departamento de Ci\^encias Exatas, Universidade Federal da Para\'{\i}ba, 58297-000 Rio Tinto, PB, Brazil}

\abstract{This work deals with charged nontopological solutions that appear in relativistic models described by a single complex scalar field in two-dimensional spacetime. We study a model which supports novel analytical configurations of the Q-ball type, that engender double exponential tails, in this sense being different from both the standard and compact Q-balls. The analytical solutions are also stable, both classically and quantum-mechanically, and the stability follows as in the case of the compact configurations.}
\maketitle
{\it {1. Introduction.}} The presence of Q-balls in relativistic models described by a single complex scalar field in $(1,1)$ spacetime dimensions was initiated in 1985 by Coleman in Ref. \cite{C} and was further investigated by several authors; see, e.g., Refs. \cite{1,2,3,4,5,6,7,8,9,10,11,12,13,14,15,16} and references therein. More recently, Q-balls have also been studied with other motivations in Refs. \cite{N0,N1,N2,N3,N4,N5,N6}. In \cite{N0}, for instance, the author considered the possibility of supermassive compact objects residing in galactic centers be giant nontopological solitons or Q-balls made of scalar fields.
In \cite{N1} the investigation focused on the linear perturbation of classically stable Q-balls solutions, and in \cite{N2} the authors investigated the limiting case of a vanishing potential term, which yielded an example of a hairy Q-ball. Moreover, in \cite{N3,N4} the authors described the existence of models which support the presence of composed soliton solutions, with  vortex and Q-ball constituents. In \cite{N5} the investigation dealt with the problem of classical stability of Q-balls, including the nonlinear evolution of classically unstable Q-balls, as well as the behaviour of Q-balls in external fields. Also, in \cite{N6} the authors studied a model described by two scalar fields to describe nontopological soliton solutions consisting of two Q-ball components possessing opposite electric charges.

In 1998, it was shown by Kusenko and Shaposhnikov in Ref. \cite{5} that Q-balls can be produced in the early universe in supersymmetric extensions of the standard model, and can contribute to dark matter.  The interesting perspective is related to the presence of Q-balls as dark matter candidate under the Affleck-Dine mechanism, which is postulated for explaining baryogenesis during the primordial expansion, after the cosmic inflation. In this line, the reviews on dark matter \cite{review} and on the origin of the matter-antimatter asymmetry \cite{review2} are of current interest. The issue in general requires the presence of scalar and gauge fields and three space dimensions, and has been further studied with several motivations. In particular, more recently a new type of charged Q-ball dark matter scenario has been suggested; see, e.g., \cite{new} and references therein. The charged solutions can carry both baryon and lepton charges, and can be stable due to the baryonic component, so they can be a viable candidate for dark matter.

In the simpler context with a single space dimension, global Q-balls were also studied by us in the recent works \cite{DB1,DB2,DB3}. There, we focused mainly on the presence of analytical solutions with distinct behavior. In \cite{DB2}, in particular, we investigated Q-balls with the novel compact profile. The analytical results that we have found in \cite{DB1,DB2,DB3} motivated us to further examine the subject to find other analytical configurations that are stable, but different from the standard and the compact Q-balls. In this sense, in the current work we continue the search for analytical configurations, but now we concentrate on the presence of solutions that are different from the standard and the compact configurations. As one knows, even though we are working in $(1,1)$ spacetime dimensions, the mandatory presence of nonlinearities and constraints complicates the investigation of stable analytical solutions of the Q-ball type.

In this work, we deal with these issues examining in the next Sec. {\it 2} the basic properties of Q-balls, accounting for the main characteristics of the standard and compact Q-balls. We then move on and in Sec. {\it 3} we introduce and solve another model, and there we also investigate the Q-ball stability. The results show the presence of exact solutions, which behave very much like the compact structures that we found in \cite{DB2}, although they  are not compact at all. In Sec. {\it 4} we end the work, adding comments and suggestions of new studies in the subject.    

{\it 2. Generalities.} 
\label{sec:gen}
In order to investigate the Q-balls in the simplest scenario, we consider the action of a single complex scalar field, $\vphi$, in $(1,1)$ Minkowski spacetime in the usual form, with the Lagrangian 
\ben
{\cal L} &=& \frac12 \partial_\mu{\bar\vphi} \partial^\mu \vphi - V(|\vphi|),
\een
with the overline on $\varphi$ denoting the complex conjugation and $V(|\vphi|)$ representing the potential. The field modulus is defined as $|\vphi| = \sqrt{{\bar\vphi}\vphi}$, and the model is invariant under rotations in the complex space of the field, $\vphi\to\vphi\, e^{i\alpha}$, with $\alpha$ real constant. One may vary the action corresponding to the above Lagrangian to get the equation of motion
\be\label{eomt}
\ddot{\vphi} - \vphi^{\prime\prime} + \frac{\vphi}{|\vphi|} V_{|\vphi|} = 0,
\ee
with the dot and the prime denoting the derivative with respect to time, $t$, and to the spatial coordinate, $x$, respectively, and $V_{|\vphi|}=dV/d|\vphi|$. To search for Q-balls we take the usual {\emph{ansatz}}
\be\label{ansatz}
\vphi(x,t)=\sigma(x)\,e^{i\omega t}.
\ee
In the above equation, $\omega$ represents an angular frequency. For simplicity, we only consider non-negative $\omega$. Notice that $|\vphi(x,t)|=\sigma(x)$, which does not depend on the time.

The presence of a phase invariance in the Lagrangian density is associated to a conserved Noether charge that has the form
\be\label{charge}
\begin{split}
	Q = \frac{1}{2i} \int_{-\infty}^\infty{dx\left({\bar\vphi}\dot{\vphi} - \vphi\dot{\bar\vphi}\right)}
	  =\omega \int_{-\infty}^\infty{dx\, \sigma^2(x)}.
\end{split}
\ee
The equation of motion \eqref{eomt} when combined with the time-dependent {\it ansatz} \eqref{ansatz} becomes 
\be\label{eom}
\sigma^{\prime\prime} = V_\sigma -\omega^2\sigma.
\ee
As usual, the boundary conditions are
\ben\label{bcond}
\sigma^\prime(0) = 0\quad\text{and}\quad\sigma(\pm\infty) \to 0.
\een
One can show that the equation of motion \eqref{eom} can be rewritten as a simpler expression
\be\label{eqeff}
\sigma^{\prime\prime} = U_\sigma,
\ee 
where $U=U(\sigma)$ acts as an effective potential for the field $\sigma$. It is given by
\be\label{veff}
U(\sigma) = V(\sigma) - \frac12\omega^2\sigma^2.
\ee
Notice that the effective potential depends explicitly on $\omega$. Therefore, one has to be careful when choosing the potential $V(|\vphi|)$ because it has to lead to a $U(\sigma)$ that allows for the presence of solutions compatible with the boundary conditions in Eq.~\eqref{bcond}. In this case, one can show that, in order to attain this compatibility, $\omega$ must be in the interval
\be\label{condomega1}
\omega_-<\omega<\omega_+,
\ee
with $\omega_+ = V_{\sigma\sigma}(0)$ and $\omega_-=\sqrt{2V(\sigma_0)/\sigma_0^2}$, where $\sigma_0$ denotes the minimum of the ratio $V(\sigma)/\sigma^2$. Here, $\omega_+$ and $\omega_-$ are the upper and lower bound for the frequency, respectively. 

Invariance over spacetime translations leads to the energy-momentum tensor
\be\label{emt}
T_{\mu\nu} = \frac12\partial_\mu{\bar\vphi}\partial_\nu\vphi + \frac12\partial_\mu\vphi\partial_\nu{\bar\vphi} - \eta_{\mu\nu}{\mathcal L}.
\ee
The component $T_{00}$ in the above equation is the energy density, which is denoted by $\epsilon$. It can be written as a sum of the kinetic, gradient and potential energy densities, $\epsilon= \epsilon_k + \epsilon_g + \epsilon_p,$
which are respectively given by
\bes\begin{align}
	\epsilon_k  = \frac12|\dot{\vphi}|^2,\quad \epsilon_g = \frac12|\vphi^{\prime}|^2 \quad\text{and}\quad \epsilon_p =  V(|\vphi|).
\end{align}
\ees
With the {\emph{ansatz}} in Eq.~\eqref{ansatz}, the energy density becomes
\be\label{edens}
\epsilon = \frac{1}{2}\omega^2\sigma^2 + \frac12{\sigma^\prime}^2 + V(\sigma).
\ee
By integrating it, we get the total energy of the Q-ball. One can also show that, by using the expression for the conserved charge in Eq.~\eqref{charge}, the kinetic energy can be written as
\be\label{ke}
E_k=\frac12 \omega Q.
\ee
Regarding the other components of the energy-momentum tensor in Eq.~\eqref{emt} with the Q-ball {\emph{ansatz}} we get $T_{01}=T_{10}=0$ and the stress
\be
T_{11} = \frac12\omega^2\sigma^2 + \frac12{\sigma^\prime}^2 - V(\sigma).
\ee
As one knows, the energy-momentum tensor is conserved, obeying the equation $\partial_\mu T^{\mu\nu}=0$. Since $T_{01}=0$, we see that $T_{11}$ cannot depend on $x$. We define the charge density as the function that is being integrated in Eq.~\eqref{charge}, $\rho_Q = \omega\sigma^2$, and this allows us to write the energy associated to Eq.~\eqref{edens} as
\be
\begin{split}
E&= \frac{\omega^2}{2}\!\int_{-\infty}^{\infty}\!\!\! dx\,\sigma^2 + \frac12\int_{-\infty}^{\infty}\!\!\! dx\,{\sigma^\prime}^2 \!\!+\! \int_{-\infty}^{\infty}\!\!\! dx\, V(\sigma)\\
&=\frac{Q^2}{2\int_{-\infty}^{\infty}\! dx\,\sigma^2} + \frac12\int_{-\infty}^{\infty}\!\!\! dx\,{\sigma^\prime}^2 \!\!+\! \int_{-\infty}^{\infty}\!\!\! dx\, V(\sigma).
\end{split}
\ee
The above expression is important because it allows us to perform variations in the energy keeping the charge fixed. At this point, we investigate how the Q-ball behaves under a rescale in the spatial coordinate, following a direction similar to the one in Refs.~\cite{lambda1,lambda2,lambda3,lambda4}. We then take $x\to z=\lambda x$, which makes $\sigma\to \sigma^{(\lambda)}(x) = \sigma(z)$, and calculate the energy of the rescaled solution using the above equation to get
\be
\begin{split}
	E^{(\lambda)} &= \frac{Q^2}{2\int_{-\infty}^{\infty}\! dx\,{\sigma^{(\lambda)}}^2} \!+ \!\frac12\!\int_{-\infty}^{\infty}\!\!\!\! dx\,{{\sigma^{(\lambda)}}^\prime}^2 \!\!+\!\! \int_{-\infty}^{\infty}\!\!\!\! dx\, V(\sigma^{(\lambda)})\\
	 &= \frac{\lambda Q^2}{2\int_{-\infty}^{\infty}\! dz\,\sigma^2(z)} \!+\! \frac{\lambda}{2}\!\int_{-\infty}^{\infty}\!\!\!\!\! dz\,\sigma^2_z(z)\! +\! \frac{1}{\lambda}\!\int_{-\infty}^{\infty}\!\!\!\!\! dz\, V(\sigma(z)),
\end{split}
\ee
where $\sigma_z(z)=d\sigma(z)/dz$. The stability against contractions and dilations in the solutions requires that $\lambda=1$ minimizes $E^{(\lambda)}$. This requirement is satisfied by the stressless condition, i.e., $T_{11}=0$. Therefore, we can write
\be\label{fo}
\frac12{\sigma^\prime}^2 = U(\sigma),
\ee
with $U(\sigma)$ being the effective potential described by Eq.~\eqref{veff}. It is straightforward to show this first-order equation is compatible with the equation of motion \eqref{eqeff}.

The stressless condition also has consequences on the energy density \eqref{edens}; it allows us to relate its contributions by
$\epsilon_p = \epsilon_k + \epsilon_g.$ The same is valid for the corresponding energies. In this case, the total energy is
$E = 2(E_k+E_g).$

We note that the stressless condition only ensures stability under contractions and dilations. However, the stability of Q-balls may be also investigated in other directions; we consider the two following possibilities \cite{15}: 
\begin{enumerate}[(i)]
\item The quantum mechanical stability, which is the stability against decay into free particles. From Eq.~\eqref{condomega1}, $\omega$ must be in a specific range of values in order to get Q-balls solutions. The Q-ball is quantum-mechanically stable if the ratio between the energy and the charge, for any $\omega$ allowed in the aforementioned range, satisfies $E/Q < \omega_+$. 

\item The classical stability, which is the one associated to small perturbations of the field. The Q-ball is classically stable if $dQ/d\omega<0$, i.e., the charge $Q$ is monotonically decreasing with $\omega$.
\end{enumerate}
There is a third type of stability, which is against fission. However, as shown in Ref.~\cite{15}, classically stable Q-balls are also stable against fission.

The form of the solution depends on the potential $V(|\vphi|)$ that one considers in the equation of motion \eqref{eom}. Below, we review the standard Q-ball, driven by the $|\vphi|^4$ potential investigated in Ref.~\cite{DB1}, and the compact Q-ball described in Ref.~\cite{DB2}, whose solution and energy density exists only in a compact space. Then, we introduce a new model, which is not compact but attains some properties of the compact structure. For simplicity, we work with dimensionless units.

{\it 2.A. Standard Q-ball.} Our first example is with the model, driven by the $|\vphi|^4$ potential
\be\label{phi4}
V(|\vphi|) = \frac12 |\vphi|^2 - \frac13|\vphi|^3 + \frac14 a\,|\vphi|^4,
\ee
with $a$ being a real and positive parameter, $a\geq0$. This model was studied in Refs.~\cite{1,2,DB1}. As one knows, due to the presence of time in the ansatz \eqref{ansatz}, this potential is influenced by the angular frequency in the equation of motion. This gives rise to an effective potential that can be calculated from Eq.~\eqref{veff}; it is given by
\be\label{vstd}
U(\sigma) = \frac12 (1-\omega^2)\sigma^2 -\frac13 \sigma^3 + \frac14 a\,\sigma^4.
\ee
One may use the first order equation \eqref{fo} to show that the solution has the form
\be\label{solstd}
\begin{aligned}
	\sigma_s(x) &= \sqrt{\frac{1-\omega^2}{2a}}\left[\tanh\left(\frac12 \sqrt{1-\omega^2}\, x + b \right) \right. \\
&\hspace{4mm}\left.-\tanh\left(\frac12 \sqrt{1-\omega^2}\, x - b \right)\right],
\end{aligned}
\ee
where $b= \text{arctanh}\left({3\sqrt{(1-\omega^2)a/2}}\right)/2$. The above expression obeys the boundary conditions \eqref{bcond} for several values of $\omega$ according to Eq.~\eqref{condomega1}. For $a\geq2/9$, we have $\omega_-=\sqrt{1-2/(9a)}$ and $\omega_+=1$. This solution is bell shaped and goes asymptotically to zero exponentially. In fact, one can show that
\be\label{large}
\sigma_s(x) \propto e^{-\sqrt{\omega^2_+-\omega^2}\,|x|}
\ee
for $x\to\pm\infty$. One can use the exact solution above to calculate the conserved charge in Eq.~\eqref{charge}, which yields to
\be\label{chargestd}
Q_s=\frac{4\omega\sqrt{1-\omega^2}}{a} \left(2b\coth(2b)-1\right).
\ee
This charge goes to infinity as $\omega\to\omega_-$ for $a>2/9$. The case $a=2/9$ is special, such that $Q_s\to0$ for $\omega\to\omega_-$. In the other side of the angular frequency range, for $\omega\to\omega_+$, the charge vanishes. We have shown in the recent investigation \cite{DB1} that the solution \eqref{solstd} exhibits quantum mechanical stability for $a>0.22268$. The classical stability is only achieved for $a>0.22540$, which is the range that ensures all of the aforementioned types of stability.

{\it 2.B. Compact Q-Balls.} A Q-ball with non-standard behavior, which engenders a compact profile, was presented in Ref.~\cite{DB2}, motivated by the work \cite{cl}. It arises with the potential
\be\label{new}
V(|\vphi|) = \frac12 |\vphi|^2 - \frac13|\vphi|^{2-1/r} + \frac14 a\,|\vphi|^{2-2/r},
\ee
where $r$ is a real parameter such that $r>2$ and $a\geq0$. In this potential, one notices the presence of fractional powers of
$|\varphi|$. As far as we know, the appearance of fractional powers in the scalar field was explored before in \cite{prl}, with the aim to generate stable kinklike configuration in arbitrary dimensions, circumventing the Derrick-Hobart scaling theorem \cite{D,H}. In particular, in the case of a single spatial dimension, one has shown in \cite{prl} that the presence of fractional power modifies the kink profile, inducing an internal structure to the field configuration. Interestingly, this kind of internal structure was found experimentally in \cite{prbrc}, in the study of domain walls in the micrometer-sized $\rm {Fe_{20}Ni_{80}}$ magnetic material in the presence of constrained geometries.

We go on and use the above potential \eqref{new} taking the usual procedure to write the effective potential \eqref{veff} in the form
\be\label{veffc1}
U(\sigma) = \frac12(1-\omega^2) \sigma^2 - \frac13\sigma^{2-1/r} + \frac14 a\,\sigma^{2-2/r}\,.
\ee
This effective potential admits a minimum at $\sigma=0,\,\forall\omega$, and a (non-minimum) zero that depends on $\omega$, at $\sigma\neq0$. So, there is a non-topological solution that connects these points. Here, the condition \eqref{condomega1} is still valid, and gives $\omega_-=\sqrt{1-2/(9a)}$ and $\omega_+=\infty$ in this case. Again, we take $a\geq 2/9$ to assure that $\omega$ is real. Then, the solutions of Eq.~\eqref{eom} are valid for any $\omega >\omega_{-}$.

One may use the first order equation \eqref{fo} to calculate the solution, which admits different forms depending on the value of $\omega$. First, we consider the case $\omega_-<\omega<1$, in which we get
\be\label{solcomp1}
\sigma_c(x) =\begin{cases}
 {\left(\frac{1-\omega^2}{2a}\right)}^{-r/2}\left[\coth\left(\frac{\sqrt{1-\omega^2}}{2r}\, x + b \right) \right. & \\
 \left.-\coth\left(\frac{\sqrt{1-\omega^2}}{2r}\, x - b \right)\right]^{-r}, & |x|\leq x_c, \\
 0, & |x|>x_c,
\end{cases}
\ee
where $x_c=2rb/\sqrt{1-\omega^2}$ delimits the compact space $-x_c\leq x\leq x_c$, in which the solution lives, and $b$ is a parameter that depends on $a$ and $\omega$ as 
\be
b=\textrm{arccoth}\left(\frac{{\sqrt{2}}+\sqrt{2-9(1-\omega^2)a}}{3\,\sqrt{a(1-\omega^2)}}\right).
\ee
The solution has a different expression for $\omega=1$, given below
\be\label{solcomp2}
\sigma_c(x) =\begin{cases}
 \left( \frac{3a}{4} -\frac{x^2}{6r^2}\right)^r, & |x|\leq x_c, \\
 0, & |x|>x_c,
\end{cases}
\ee
where $x_c$ now changes to $x_c=3r\sqrt{a/2}$.

For $\omega>1$, one can show that $\sigma(x)$ is written by
\be\label{solcomp3}
\sigma_c(x) =\begin{cases}
 {\left(\frac{\omega^2-1}{2a}\right)}^{-r/2}\left[\cot\left(\frac{\sqrt{\omega^2-1}}{2r}\, x + d \right) \right. & \\
 \left.-\cot\left(\frac{\sqrt{\omega^2-1}}{2r}\, x - d \right)\right]^{-r}, & |x|\leq x_c, \\
 0, & |x|>x_c,
\end{cases}
\ee
where $x_c=2rd/\sqrt{\omega^2-1}$ and $d$ is given by 
\be
d=\text{arccot}\left(\frac{\sqrt{2}+\sqrt{2+9(\omega^2-1)a}}{3\sqrt{a(\omega^2-1)}}\right).
\ee
Regardless the value of $\omega$, the solution vanish outside a compact interval of the real line. This is a different behavior from the solution of the standard case described in Eq.~\eqref{solstd}, which only vanish asymptotically, going to zero with an exponential tail. Therefore, compact Q-balls do not have a tail, so they seem to behave as hard charged balls.

An interesting fact is that, even though the solution has different expressions, depending on the value of the angular frequency, as in Eqs.~\eqref{solcomp1}, \eqref{solcomp2} and \eqref{solcomp3}, one can calculate the conserved charge in Eq.~\eqref{charge} to show that it obeys the following single expression
\be\begin{aligned}\label{chargecomp}
&Q_c=\frac{2\,r\,\omega\sqrt{2\pi}\,3^{2r+1}a^{2r+1/2}}{\left( 2 \left(1+\sqrt{1-9(1-\omega^2)a/2}\right)\right)^{2r+1}}\frac{\Gamma(2r+1)}{\Gamma\left(2r+\frac32\right)} \\
& _2F_1\left(\frac12,2r+1;2r+\frac32;\frac{9(1-\omega^2)a/2}{\left(1+\sqrt{1-9(1-\omega^2)a/2}\right)^2}\right), \\
\end{aligned}
\ee
where $\Gamma(z)$ is the Gamma function and $_2F_1\left(a,b;c;z\right)$ is Hypergeometric function. Notice that the above expression depends on $r$, $a$ and $\omega$. Similarly to what happens with the charge in Eq.~\eqref{chargestd} for the standard case, we have $Q_c\to\infty$ in the limit $\omega\to\omega_-$ for $a>2/9$, and $Q_c\to0$ if $\omega\to\omega_-$ for $a=2/9$. The limit $\omega\to\omega_+$ leads to null charge; this does not depend on the $a$ chosen. 

Regarding the stability of the compact Q-balls, it was shown in Ref.~\cite{DB2} that they are quantum mechanically stable, because $\omega_+$ is infinity. This means that the inequality $E_c/Q_c<\omega_+=\infty$ is satisfied for $a\geq2/9$. Therefore, compact Q-balls never decay into free particles, for $r>2$. On the other hand, the classical stability is not ensured and must be investigated carefully. In the case of $r=3$, it was shown in \cite{DB2} that the compact Q-ball is classically stable for $a>0.22249$, in which the charge is monotonically decreasing with $\omega$. 

\begin{figure}[b!]
\centering
\includegraphics[width=5cm]{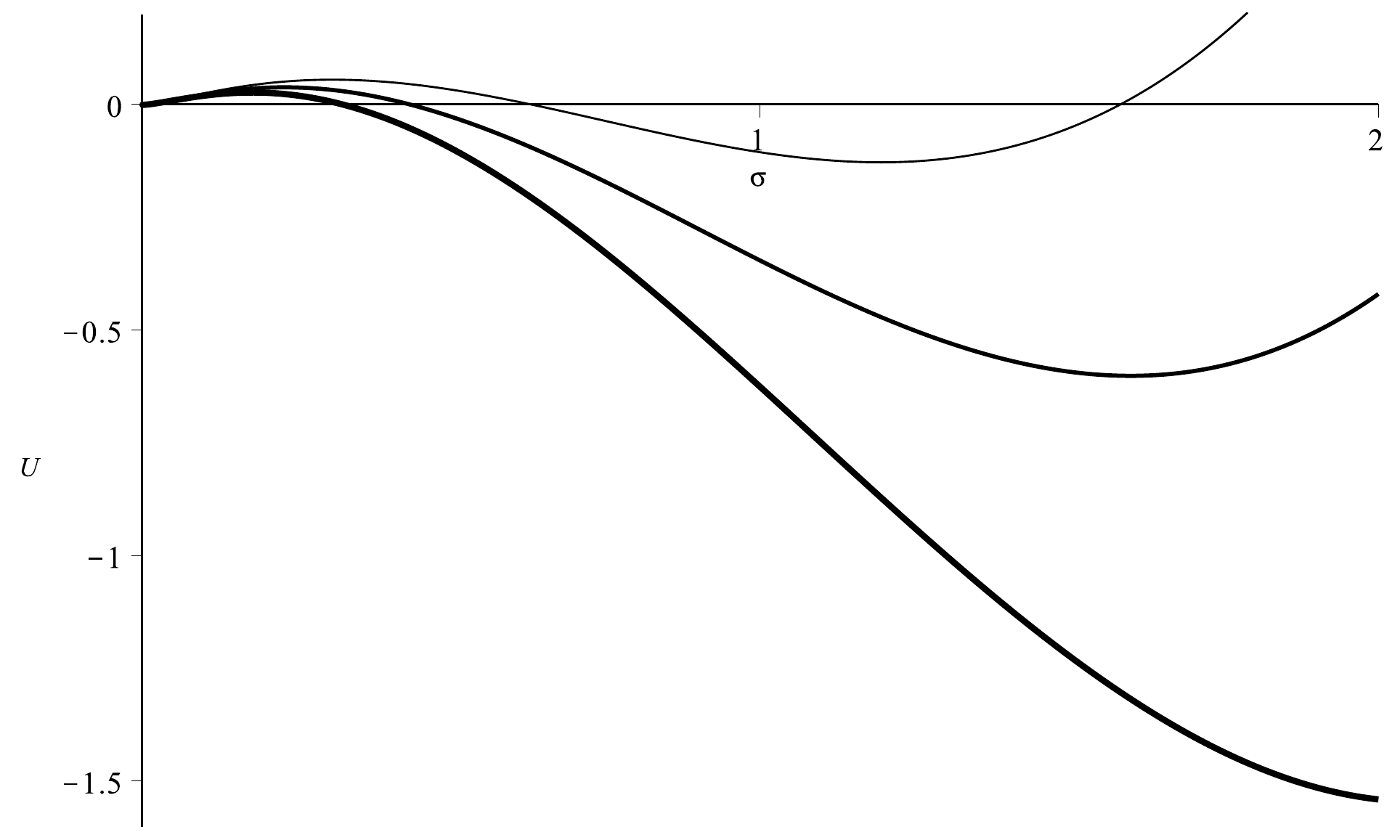}
\caption{The effective potential in Eq.~\eqref{vefffcomp} depicted for $a=1$, and $\omega=1.1,\,1.3$ and $1.5$. The thickness of the lines increases with $\omega$.}
\label{fig1}
\end{figure}

{\it 3. Novel structures.} In this section, we get inspiration from the lumplike structures recently introduced in Ref.~\cite{dexplump} and present a new model which supports Q-balls with new features that make them special. We consider the potential
\be\label{vfcomp}
V(|\vphi|) = \frac12|\vphi|^2\left(a^2+\ln^2|\vphi|\right).
\ee
Here, $a$ is a real non-negative parameter that is squared for convenience. Potentials with a logarithmic term appeared before in Refs.~\cite{log1,log2}; in particular in \cite{log2} the author investigated the formation of Q-balls in the presence of the so-called thermal logarithmic potential. The result indicates that the growth of the field fluctuations is fast enough to create Q-balls, despite the shrinking instability which is due to the decreasing temperature, etc; see also Ref. \cite{log3} for other results related to \cite{log2}. We further notice the work \cite{log4}, in which the thermal logarithmic potential is also investigated, but now with variational estimation instead of lattice simulation. The study led to analytical results on the Q-balls properties such as radio, energy and stability, without the need to solve the nonlinear field equation numerically.

As usual in the search for Q-balls, the above potential \eqref{vfcomp} is influenced by the contribution of the angular frequency in the equation of motion. This gives rise to an effective potential -- see Eq.~\eqref{veff} -- that has the form
\be\label{vefffcomp}
U(\sigma) = \frac12\sigma^2\left(a^2-\omega^2+\ln^2\sigma\right).
\ee
This effective potential admits a minimum at $\sigma=0$ and a neighbor zero at $\sigma\neq0$. As usual, the allowed range for $\omega$ is obtained through Eq.~\eqref{condomega1}. Here, we get $\omega_-=a$ and $\omega_+=\infty$. It is worth commenting, though, that $\omega_+$ is infinite as in the compact Q-ball for a different reason: this happens here due to the presence of the logarithmic term in the above potential, not because of the unusual power that appeared in the potential \eqref{veffc1}. In Fig.~\ref{fig1}, we display the behavior of the above effective potential for several values of the angular frequencies and $a=1$. We can see that, as $\omega$ increases, the zero of the potential approaches to the minimum at the origin. Meanwhile, the minimum outside the interval where the solution exists goes deeper and farther.

\begin{figure}[h!]
\centering
\includegraphics[width=4cm]{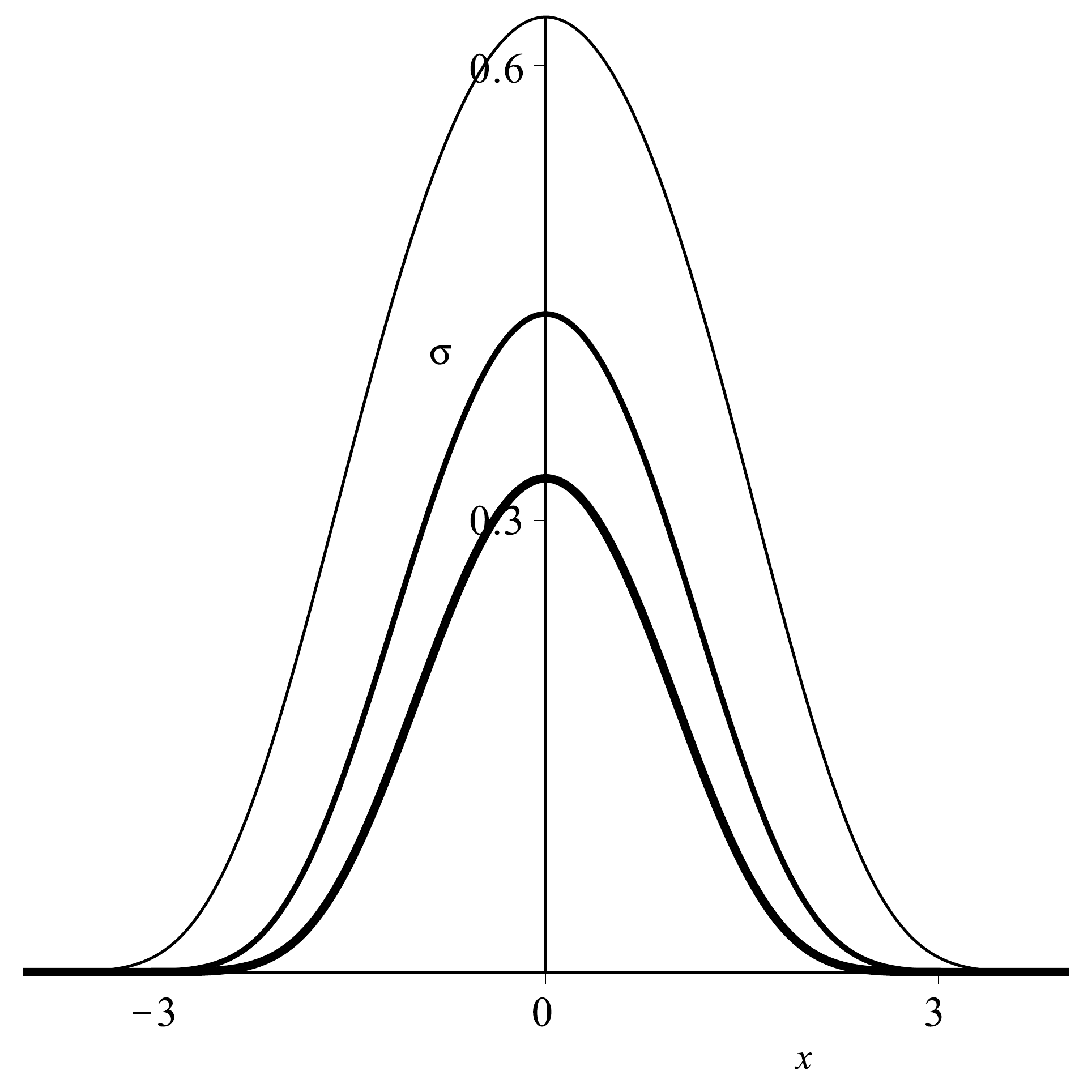}
\includegraphics[width=4cm]{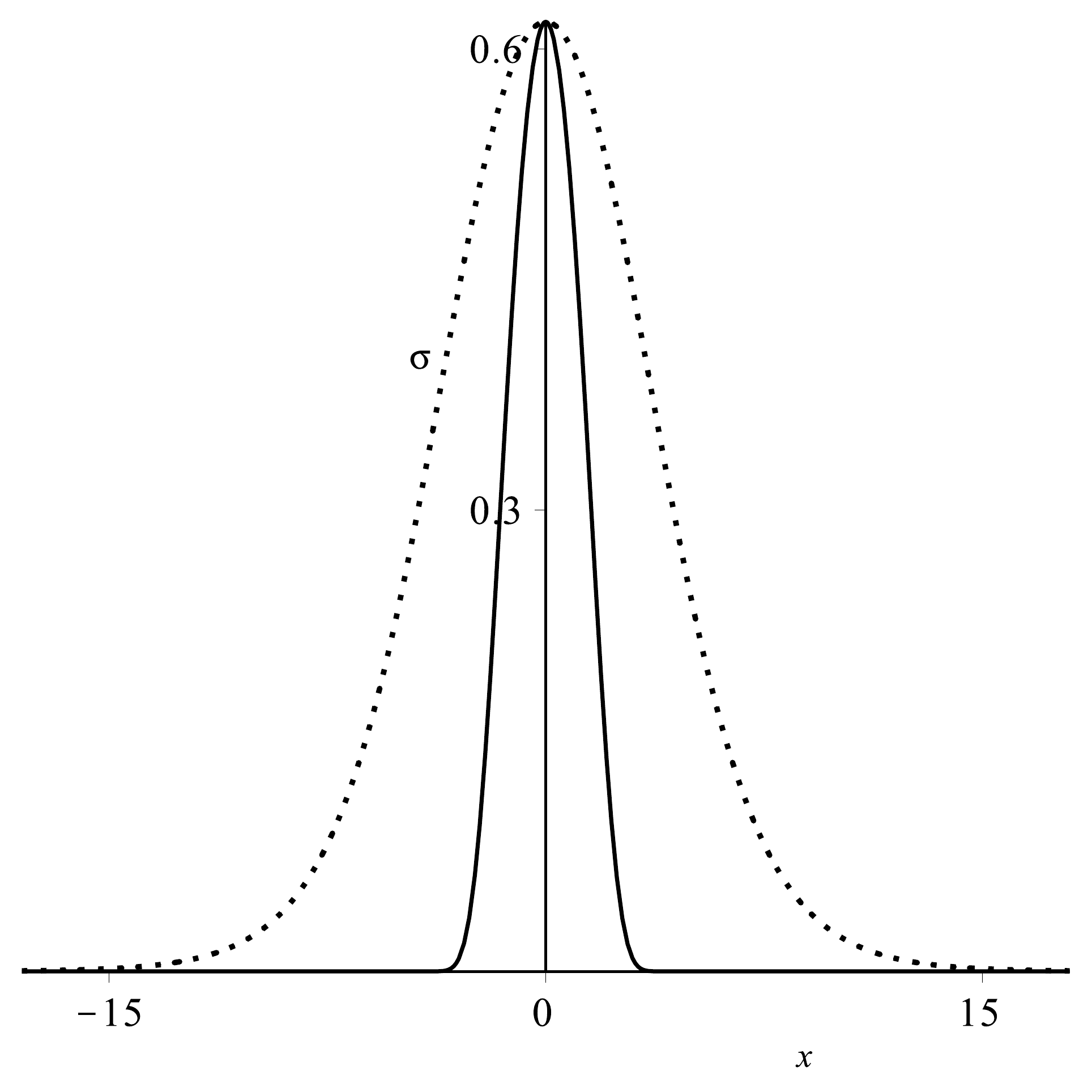}
\caption{(Left panel) The solution \eqref{solfcomp} depicted for $a=1$, and $\omega=1.1,\,1.3$ and $1.5$, with the thickness of the lines increasing with $\omega$. (Right panel) The standard solution \eqref{solstd} (dotted line) and the quasi-compact solution \eqref{solfcomp} (solid line) are shown together, for comparison.}
\label{fig2}
\end{figure}

According to the boundary conditions in Eq.~\eqref{bcond} and the first order equations \eqref{fo}, the zero of the potential which is in the neighborhood of the minimum of the potential gives the amplitude of the solution, which we denote by $A$. In this case, one can show that
$A = \exp{\bigl(-\sqrt{(\omega^2-a^2)}\bigr)}$. Thus, the amplitude of the solution starts at the maximum value $A_{\max}=1$, in the limit
$\omega\to a$, and gets smaller as $\omega$ increases, as expected from the behavior of the effective potential \eqref{vfcomp}. To calculate the solution, we use the first order equation \eqref{fo}, which becomes
\be
{\sigma^\prime}^2 = \sigma^2\left(a^2-\omega^2+\ln^2\sigma\right).
\ee
It admits an analytical solution satisfying the boundary conditions \eqref{bcond}, with the form
\be\label{solfcomp}
\sigma_q(x) = e^{-\sqrt{\omega^2-a^2}\cosh(x)}.
\ee
This solution presents a double exponential tail, i.e., it decays asymptotically as 
\be
\sigma_q\propto e^{-\sqrt{\omega^2-a^2}\,e^{|x|}},
\ee
for $x\to\pm\infty$. This solution is depicted in the left panel in Fig. \ref{fig2} and, even though it vanishes faster than the standard solution in Eq.~\eqref{solstd} as $x$ increases, see Eq. \eqref{large}, it is not compact as the solutions in Eqs.~\eqref{solcomp1}, \eqref{solcomp2} and \eqref{solcomp3}, which do not exhibit a tail. 

The above solution represents a quasi-compact Q-ball, because of the decay as the exponential of an exponential, engendering a tail suppression which is much stronger than the standard Q-balls. In order to further emphasize this characteristic, we also display in the right panel in Fig. \ref{fig2} the standard and the quasi-compact solutions together. They are shown with the same amplitude and the same $\omega$, but with distinct values of $a$, since $a$ in the standard model controls the fourth-order power of the scalar field, and in the quasi-compact model it has a very different meaning, contributing to the second-order power of the scalar field. In the right panel in Fig. \ref{fig2} we used $\omega=0.85$, and $a=0.70364$ and $a=0.7$, for the standard and the
quasi-compact solutions, respectively.

\begin{figure}[t!]
\centering
\includegraphics[width=4cm]{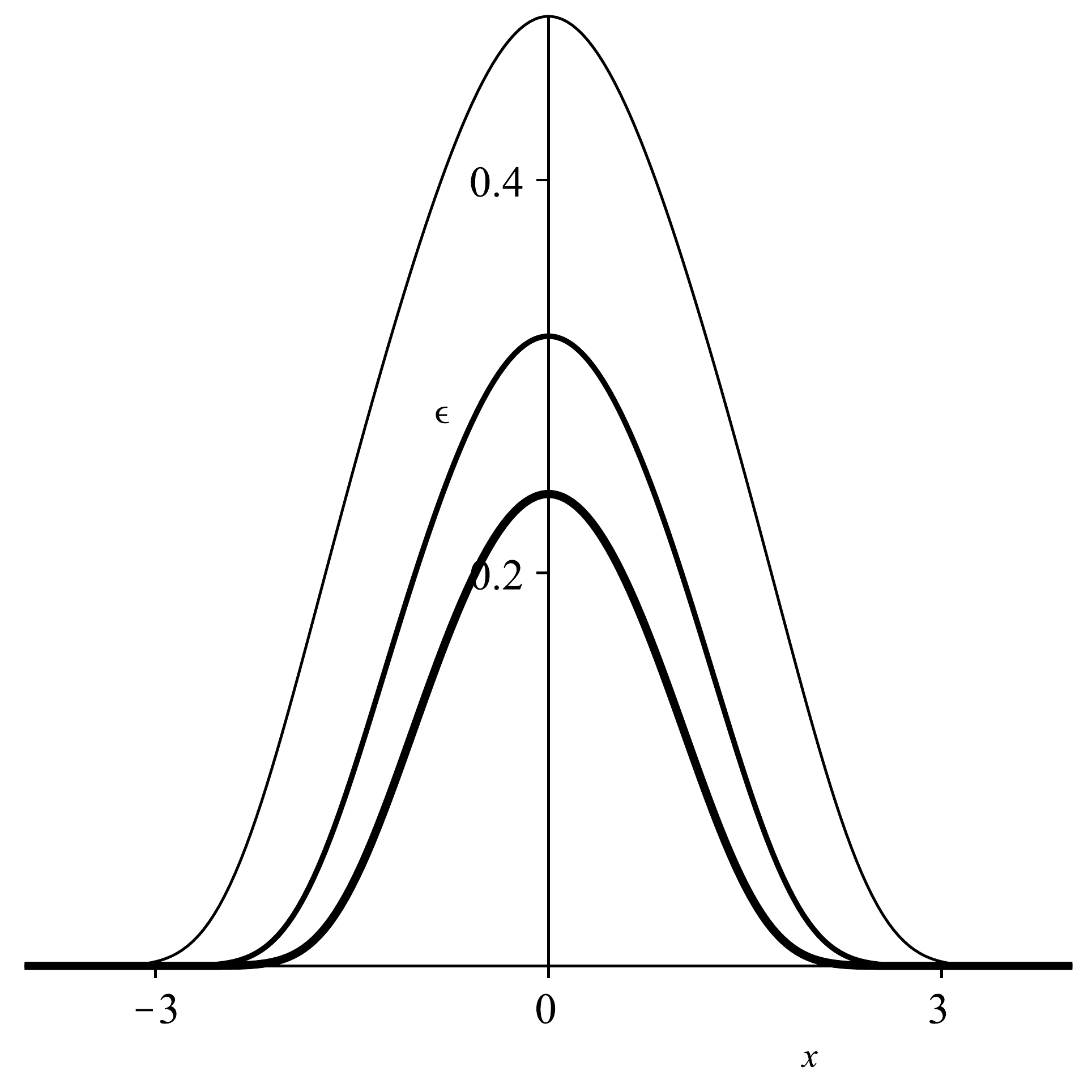}
\includegraphics[width=4cm]{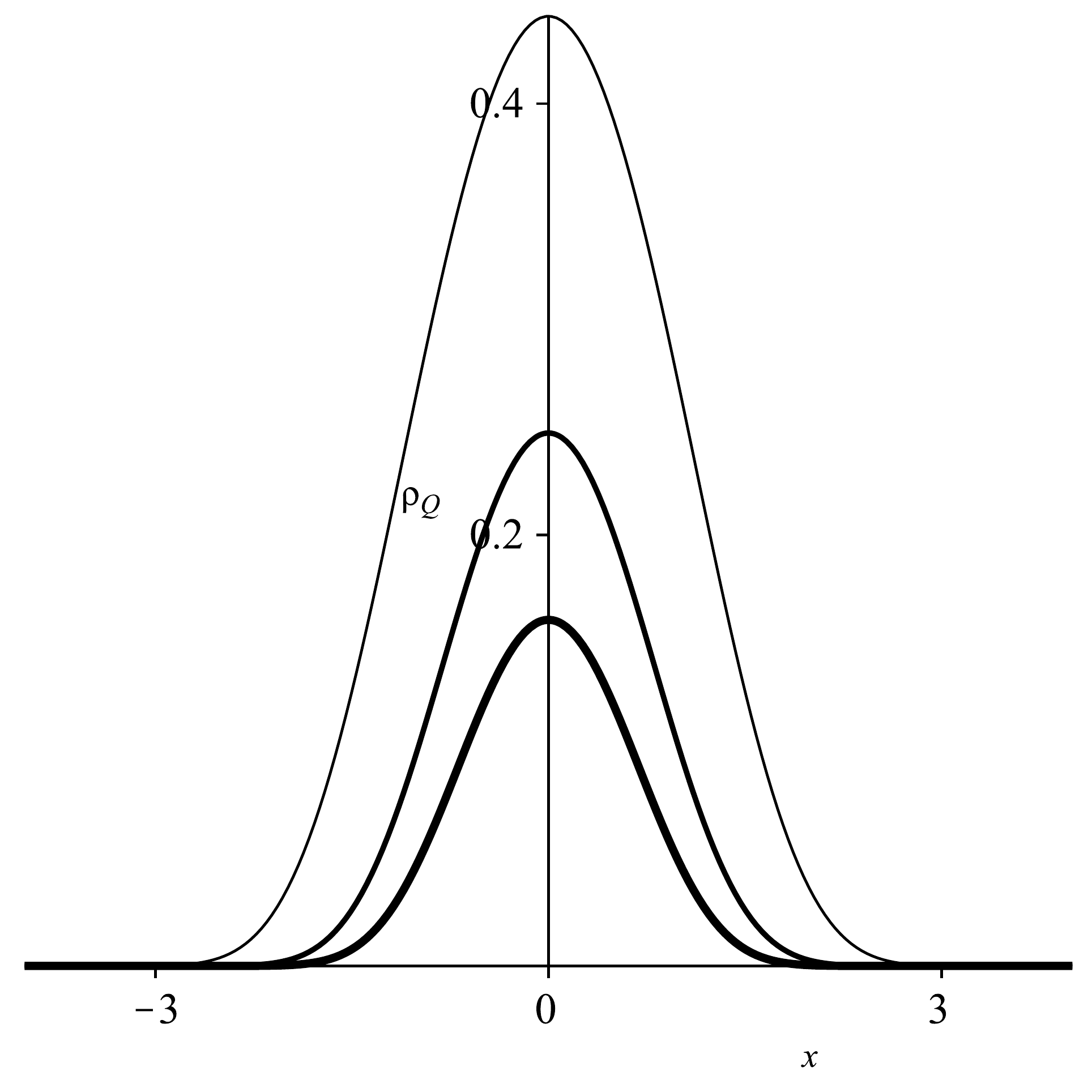}
\caption{The energy density in Eq.~\eqref{edensfcomp} (left) and the charge density in Eq.~\eqref{qdensfcomp} (right) depicted for $a=1$, and $\omega=1.1,\,1.3$ and $1.5$. The thickness of the lines increases with $\omega$.}
\label{fig3}
\end{figure}

We can use the analytical solution given above to calculate the energy density in Eq.~\eqref{edens}, which leads to
\be\label{edensfcomp}
\epsilon_q = \left(\omega^2\cosh^2(x)-a^2\sinh^2(x)\right)e^{-2\sqrt{\omega^2-a^2}\cosh(x)}.
\ee
We are also able to calculate the charge density analytically; it is $\rho_{Q_q}(x) = \omega\sigma_q^2(x)$, which is the function that is being integrated in Eq.~\eqref{charge}. It has the form
\be\label{qdensfcomp}
\rho_{Q_q} = \omega\, e^{-2\sqrt{\omega^2-a^2}\cosh(x)}.
\ee
In Fig.~\ref{fig3}, we display the energy density in Eq.~\eqref{edensfcomp} and the charge density above. Notice that they are both bell-shaped and present a double exponential tail. So, they behave similarly to the solution \eqref{solfcomp}.

\begin{figure}[t!]
\centering
\includegraphics[width=4cm]{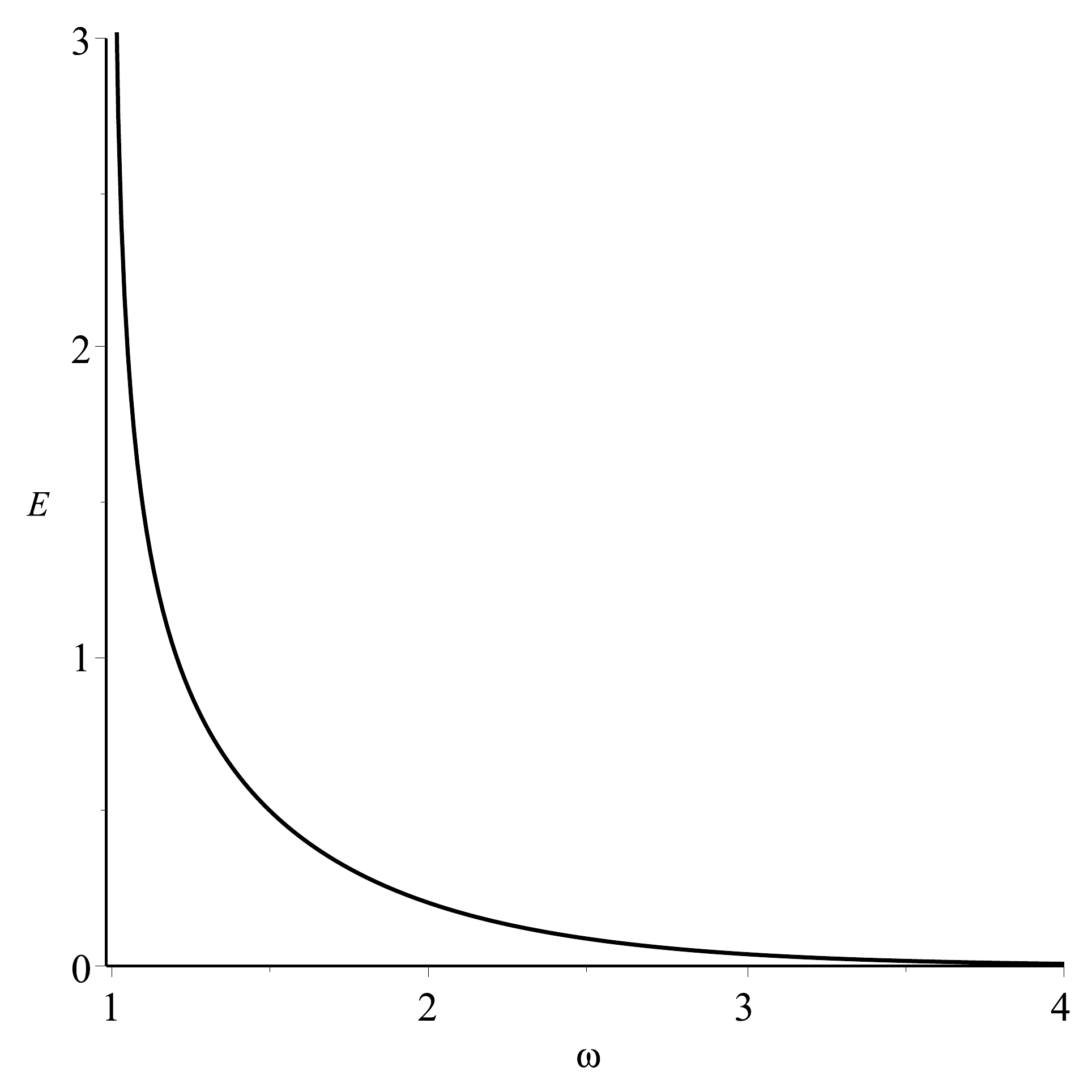}
\includegraphics[width=4cm]{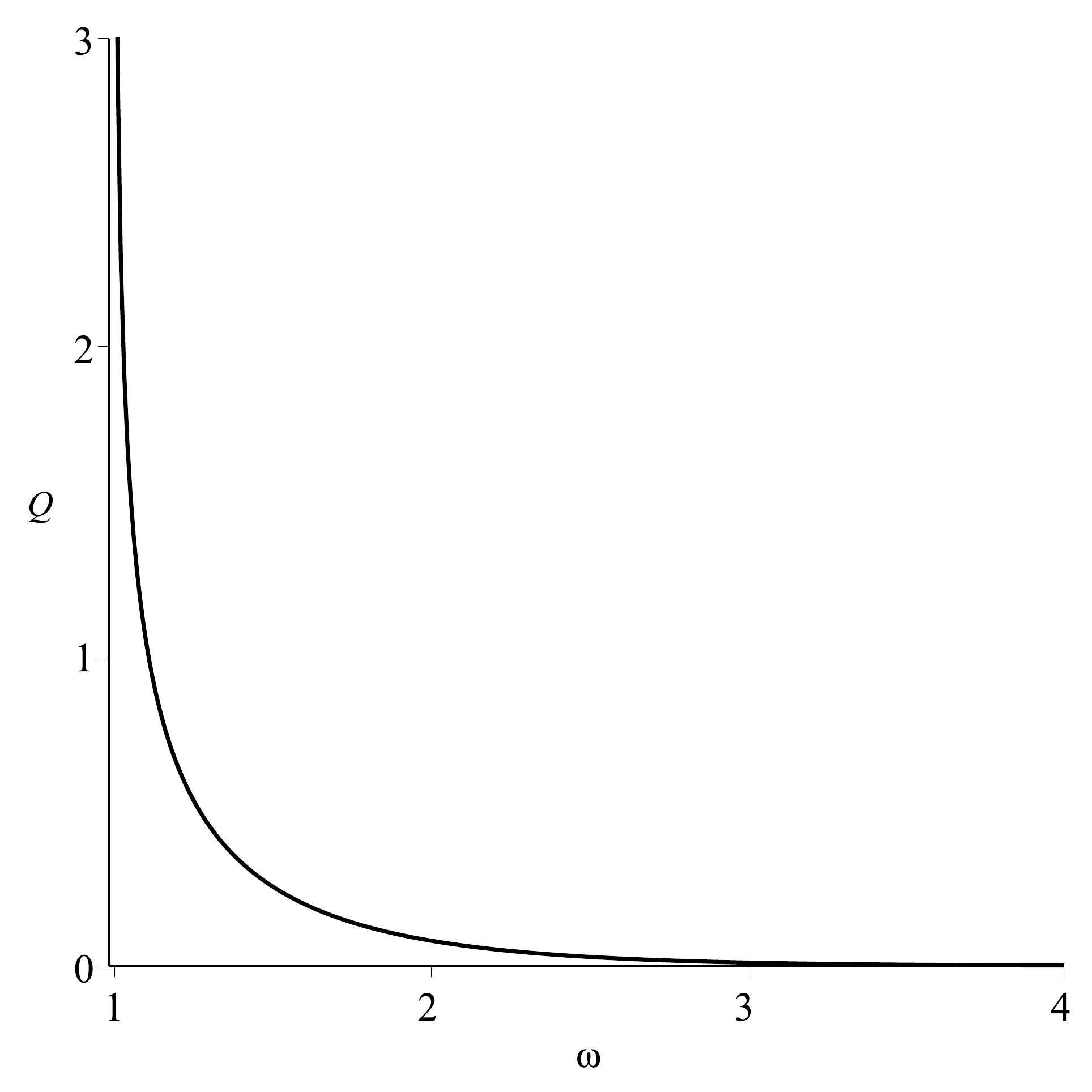}
\caption{The energy (left) and charge (right) associated to Eqs.~\eqref{edensfcomp} and Eq.~\eqref{qdensfcomp} displayed as functions of $\omega$ for $a=1$.}
\label{fig4}
\end{figure}

One may integrate the quantities in Eqs.~\eqref{edensfcomp} and \eqref{qdensfcomp} all over the space to calculate the energy $E$ and the charge $Q$ of the Q-ball, respectively. Unfortunately, we have been unable to find the analytical expressions for them, so we must proceed using numerical methods. Notice that they depend on the parameters $a$ and $\omega$, with the latter restricted by the range informed in Eq.~\eqref{condomega1}. To illustrate their behavior, we plot them for $a=1$, as a function of $\omega$, in Fig.~\ref{fig4}.

We now focus on the stability of the new Q-balls present in the model described by the potential \eqref{vfcomp}. As we discussed in Sec.~\ref{sec:gen}, both the energy and charge of the solutions play an important role in the stability of the Q-ball. As we commented before, the ratio $E/Q$ is associated to the quantum mechanical stability of the Q-ball, which is related to the stability against decay into free particles. On the other hand, the behavior of the charge $Q$ with respect to $\omega$ dictates the classical stability of the Q-ball, against small fluctuations in the field.

In order to investigate the quantum mechanical stability, we must remember that $\omega_+=\infty$. So, the condition $E_q/Q_q<\omega_+$ is always satisfied and the Q-ball is quantum mechanically stable regardless the value of $a$. This feature is the same of the compact Q-ball. Here, however, the Q-ball is not compact: it presents a tail that decays extremely fast, but not enough to become compact. So, we have found a Q-ball that presents a double exponential tail whose qualitative behavior of the quantum mechanical stability is the same of the compact Q-ball. These new structures never decay into free particles, and this is another motivation to call them quasi-compact Q-balls.

\begin{figure}[t!]
\centering
\includegraphics[width=4.2cm]{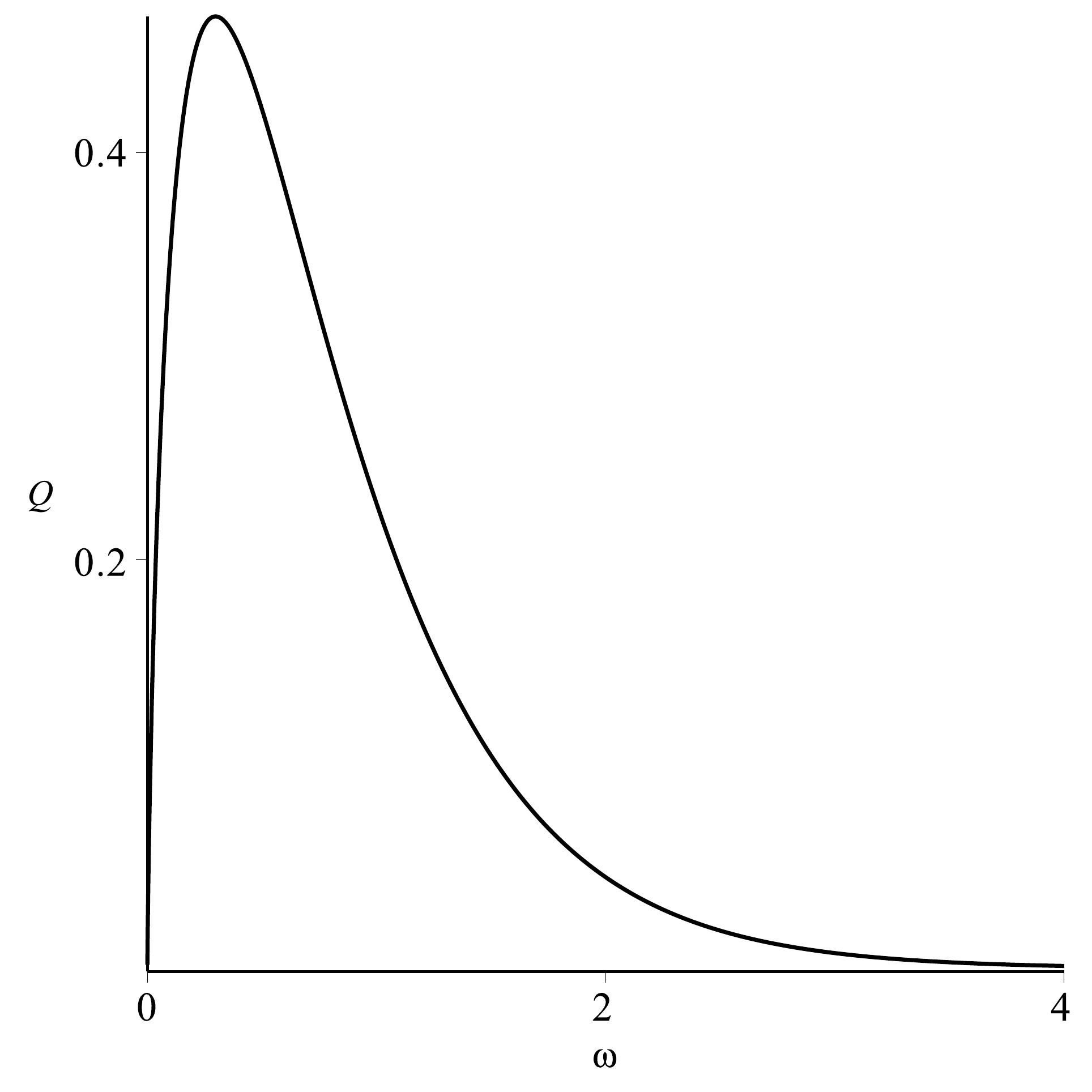}
\includegraphics[width=4.2cm]{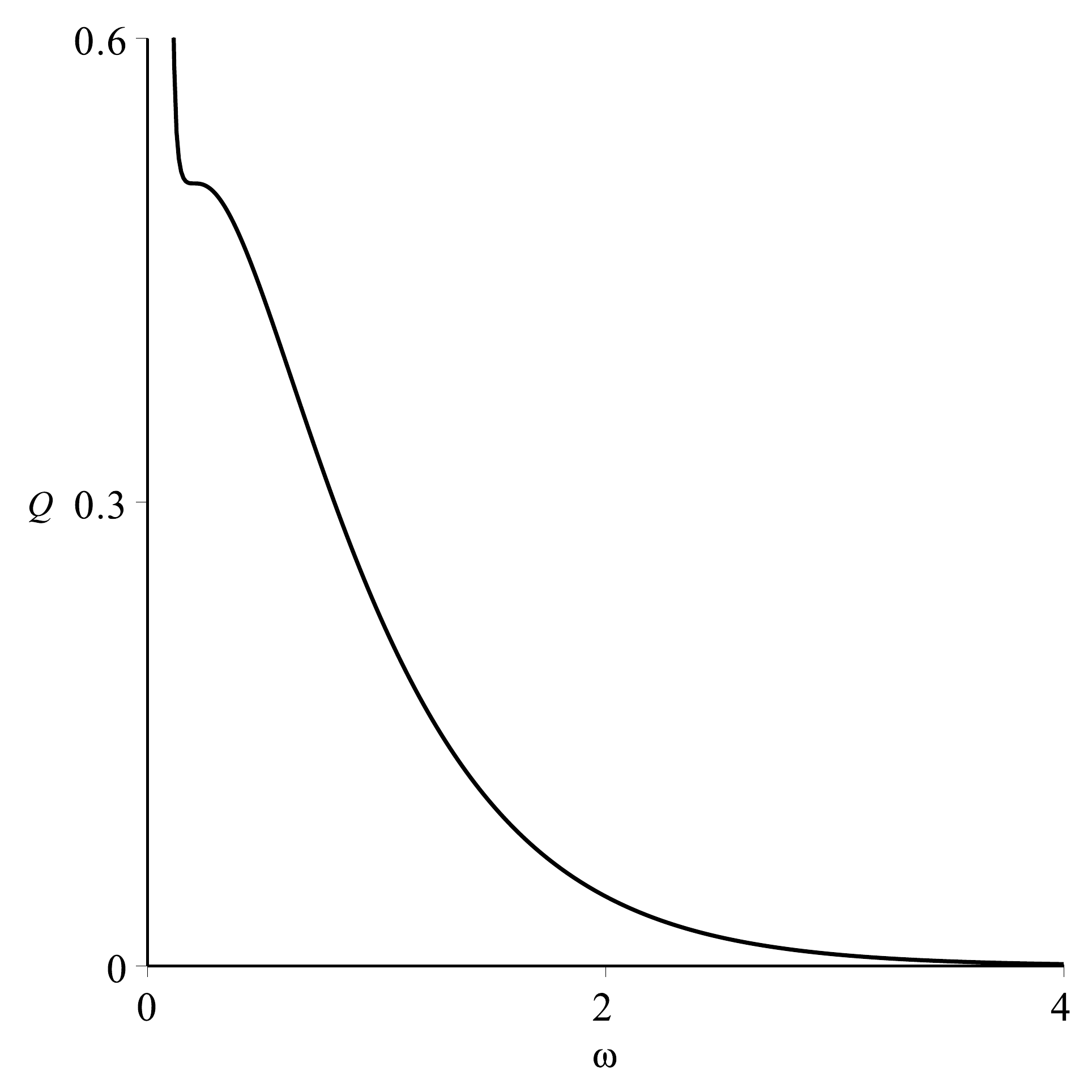}
\caption{The charge associated to Eq.~\eqref{qdensfcomp} as function of $\omega$ for $a=0$ 
(left) and $a=0.10731$ (right).}
\label{fig5}
\end{figure}

The other type of stability that we have to study is the classical one. To do that, we plot the charge for $a=0$ in terms of $\omega$ in Fig.~\ref{fig5}. In this case, the charge presents a hole near $\omega=0$; it is not monotonically decreasing with $\omega$, which makes the Q-ball unstable under small fluctuations. We continue the investigation increasing $a$ until we arrive at $a=0.10731$, which leads to an inflection point around $\omega_{\textrm{inflec}}\approx0.21$ and $Q_{\textrm{inflec}}\approx 0.5061$. This means that, for this value of $a$, we have $dQ/d\omega\leq0$. Therefore, for $a>0.10731$, the quasi-compact Q-ball is classically stable. Since the quantum-mechanical stability is attained for $a>0$, we conclude this new structure is stable for $a>0.10731$, both classically and quantum mechanically.

{\it 4. Conclusion.} We have introduced and investigated a new model of Q-balls, whose solution presents a double exponential tail. In this way, it is different from both the standard and the compact Q-balls. The associated potential presents infinite second derivative at the minimum, but this does not make the solution compact, as it happens in the model investigated in Ref.~\cite{DB2}. Nevertheless, it leads to
$\omega_+=\infty$, which is the upper limit for the angular frequency. This property is interesting, because it ensures the quantum mechanical stability as in the compact Q-ball. So, we were able to introduce Q-balls that are not compact but present the same quantum mechanical stability of the compact solutions found in \cite{DB2}, so we called them quasi-compact Q-balls. We also investigated their classical stability and concluded that they are classically stable for $a>0.10731$.

The quasi-compact Q-ball engenders the feature of having a tail that decays as the exponential of an exponential. This is novel behavior that makes the solution very different from the standard Q-ball, with matter much more concentrated around its center. This behavior is welcome, and may open new possibilities of applications of current interest. We can, for instance, think of it as a new candidate to model dark matter; see, e.g., \cite{5,new,review}. In particular, one notes in \cite{new} the presence of a logarithmic potential in the Affleck-Dine scalar field, which is due to local gauge effects which appear in the minimal gauge mediation model \cite{mediate}. Evidently, in more realistic models we have to study how the couplings with other fields act to change the features we found above, in the simple model with a single complex scalar field. To go further on this direction, one should add gauge fields and enlarge space to accommodate three spatial dimensions. Another perspective concerns the addition of gravity, since it may also act to change the profile of the solution and spoil the quasi-compact behavior found above. Research in this direction is welcome, and may follows the lines of works on gravitating monopoles \cite{M,M1} and Q-balls \cite{GQ}.

The profile of the quasi-compact Q-ball, the possibility of thinking of compact objects as Q-balls made of scalar fields \cite{N0} and the perspective of using it as dark matter candidate motivate new studies and we are now examining the extension of the above results to the $(3,1)$ dimensional scenario, with the addition of gauge and gravitational fields; see, e.g., Ref. \cite{local} for the interesting approach based on variational estimation. Other possibilities concern the two-dimensional vortex and Q-ball system described in \cite{N4} and the one-dimensional soliton system of gauged Q-ball and anti-Q-ball presented in \cite{N6}, focusing on the construction of new solutions, with the quasi-compact profile described in this work. We can also follow the recent works \cite{22,33} to enhance the $U(1)$ symmetry in a way capable of adding internal structures to the nontopological solutions. These and others related issues are currently under consideration, and we hope to report on them in the near future.\\

\acknowledgements
{The authors are grateful to Conselho Nacional de Desenvolvimento Cient\'\i fico e Tecnol\'ogico (CNPq) for partial financial support.}

\end{document}